# Phonon dynamics in the layered negative thermal expansion compounds $Cu_xNi_{2-x}(CN)_4$

Stella d'Ambrumenil[1,2], Mohamed Zbiri[1*], Ann M. Chippindale[2], Simon J. Hibble[3]

[1]Institut Laue-Langevin, 71 avenue des Martyrs, Grenoble Cedex 9, 38042, France.
[2]Department of Chemistry, University of Reading, Whiteknights, Reading, RG6 6AD, United Kingdom.
[3]Chemistry Teaching Laboratory, Department of Chemistry, University of Oxford, South Parks Road, Oxford, OX1 3PS, United Kingdom.

*zbiri@ill.fr

ABSTRACT

This study explores the relationship between phonon dynamics and negative thermal expansion (NTE) in $Cu_xNi_{2-x}(CN)_4$. The partial replacement of nickel (II) by copper (II) in $Ni(CN)_2$ leads to a line phase, $CuNi(CN)_4$ ($x = 1$), and a solid solution, $Cu_xNi_{2-x}(CN)_4$ ($0 \leq x \leq 0.5$). $CuNi(CN)_4$ adopts a layered structure related to that of $Ni(CN)_2$ ($x = 0$), and interestingly exhibits 2D NTE which is ~ 1.5 times larger. Inelastic neutron scattering (INS) measurements combined with first principles lattice dynamical calculations provide insights into the effect of $Cu^{2+}$ on the underlying mechanisms behind the anomalous thermal behavior in all the $Cu_xNi_{2-x}(CN)_4$ compounds. The solid solutions are presently reported to also show 2D NTE. The INS results highlight that as the $Cu^{2+}$ content increases in $Cu_xNi_{2-x}(CN)_4$, large shifts to lower energies are observed in modes consisting of localized in- and out-of-plane librational motions of the CN ligand, which contribute to the NTE in $CuNi(CN)_4$. Mode Grüneisen parameters calculated for $CuNi(CN)_4$ show that acoustic and low-energy optic modes contribute the most to the NTE, as previously shown in $Ni(CN)_2$. However, mode eigenvectors reveal a large deformation of the $[CuN_4]$ units compared to the $[NiC_4]$ units, resulting in phonon modes not found in $Ni(CN)_2$, whose NTE-driving phonons consist predominately of rigid-unit modes. The deformations in $CuNi(CN)_4$ arise because the $d^9$ square-planar center is easier to deform than the $d^8$ one, resulting in a greater range of out-of-plane motions for the adjoining ligands.

INTRODUCTION

Many layered compounds exhibit interesting physical properties arising directly from their 2D nature. Metallic films, magnetic layers, 2D conductors and superconductors are the basis of thin-film technology [1]. The ability of layered compounds to form intercalated systems has also resulted in many electrochemical and catalytic uses [2]. In some cases, 2-dimensional compounds exhibit the abnormal property of negative thermal expansion (NTE), which has potential uses in applications requiring a targeted thermal response, and in composite materials requiring overall zero thermal expansion. Several transition-metal cyanides exhibit NTE and many investigations have been carried out to understand the atomistic mechanisms behind the abnormal thermal behavior. Examples include $Zn(CN)_2$ and $Cd(CN)_2$ [3], which exhibit 3D isotropic NTE along with high-temperature CuCN, AgCN and AuCN, which exhibit 1D NTE [4]. More recently, the behaviors of mixed-metal cyanides, such as $Cu_xAg_{1-x}CN$, $ZnAu_2(CN)_4$, $ZnNi(CN)_2$ and $Ag_3Co(CN)_6$, as well as many other Prussian blue analogues, have been studied [5-9].

$Ni(CN)_2$ is an example of a well-studied layered transition-metal cyanide exhibiting 2D NTE [10,12-14]. It has a structure consisting of sheets of square-planar [Ni(C/N)$_4$] units, which lack long-range stacking order. The compound has thermal expansion coefficients, $α_a$ of $-6.5 \times 10^{-6}$ K$^{-1}$ and $α_c$ of $+69 \times 10^{-6}$ K$^{-1}$, where $a, b$ is the in-plane lattice parameter and $c$ is normal to the layers. These lead to an overall volume expansion coefficient, $α_V$, of $+48 \times 10^{-6}$ K$^{-1}$. Replacing half of the $Ni^{2+}$ ions with $Cu^{2+}$ to form $CuNi(CN)_4$ results in an isostructural compound with a smaller interlayer separation and more pronounced 2D NTE ($α_a = -9.7(8) \times 10^{-6}$ K$^{-1}$) [11]. The $α_c$ and $α_V$ coefficients are also correspondingly larger compared to those of $Ni(CN)_2$, and their values increase with temperature [11]. The metal atoms alternate within the sheets as in a checkerboard and, unlike in $Ni(CN)_2$, in which the cyanide ligands show 'head-to-tail' disorder, the cyanide ligands in $CuNi(CN)_4$ are completely ordered with the carbon end bonding to Ni [11]. A single sheet of $CuNi(CN)_4$ is shown in Figure 1. The mechanism behind the observed NTE in the two compounds is key to understanding how the addition of Cu enhances the phenomenon.

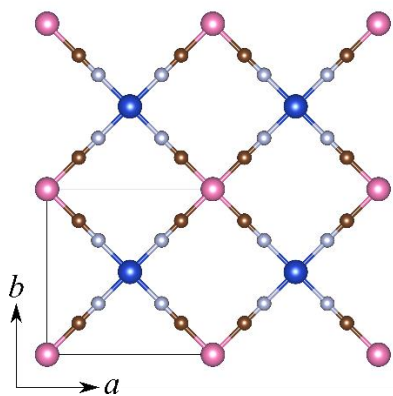

Figure 1: Schematic illustration of a single sheet of $CuNi(CN)_4$. Key: Cu, blue, Ni, pink, C, brown, N, grey. The unit cell outlined has lattice parameters $a = b = 6.99$ Å at 15 K [11]. Refined values of the lattice parameters from x-ray diffraction (XRD) and neutron pair distribution function (PDF) analysis, along with relaxed values using different DFT schemes are shown in Figure 7.

The first insight into the mechanism of 2D NTE in Ni(CN)$_2$ was found by reverse Monte-Carlo (RMC) fitting of total neutron diffraction data by Goodwin *et al.* [12]. Results yielded five dispersionless phonon modes below 4.1 meV involving rigid rotations and translations of the [Ni(C/N)$_4$] units, four of which forced the C and N atoms out of plane. The four vibrations produce a rippling effect of the layers, which has the net effect of bringing the Ni atoms closer together, whilst at the same time pushing neighboring layers further apart.

Inelastic neutron scattering (INS) provided a direct way of probing phonon dynamics in Ni(CN)$_2$ [13]. Measurements were combined with *ab initio* DFT calculations carried out using a single nickel-cyanide layer, neglecting the 'head-to-tail' disorder of the cyanide ligands. The calculated phonon density of states accurately reproduced the measured phonon bands, and the calculated value of $\alpha_a$ was comparable to that extracted from x-ray diffraction experiments [13]. The calculated mode Grüneisen parameters revealed the acoustic and first three optic modes at ~ 12 meV (a higher energy than from the RMC) as the primary source of the NTE. Normal modes analysis showed that these modes also consist of in- and out-of-plane translational and rotational motions of the [Ni(C/N)$_4$] rigid units. Furthermore, the modes which make the largest contributions at the zone boundaries consist only of out-of-plane motions. INS conducted on Ni(CN)$_2$ at variable pressure revealed that at high pressure, low-energy modes shift to higher energy [14]. This is because as the interlayer separation decreases, out-of-plane motions become more difficult.

The present study is focused on understanding the atomistic mechanisms of CuNi(CN)$_4$ given that it has a higher 2D NTE coefficient than Ni(CN)$_2$, despite having a smaller interlayer separation. We have carried out temperature-dependent INS measurements, underpinned by *ab initio* DFT calculations, on CuNi(CN)$_4$ to gain insights into how the addition of Cu$^{2+}$ affects its dynamics and thermal expansion behavior. We have also extended our INS measurements to probe the phonon spectra of Cu$_x$Ni$_{2-x}$(CN)$_4$ with compositions other than $x = 1$, namely $x = 0.1, 0.25, 0.33$ and $0.5$, which from part of a solid solution (when $0 \leq x \leq 0.5$). These materials also have layered structures [11], although it is unlikely that the layers are planar. Here, we demonstrate that they also show 2D NTE. These solid solutions deserve a dedicated work due to their structural complexity for modelling and calculations, as together with CuNi(CN)$_4$ and Ni(CN)$_2$, they form a series with interesting trends.

EXPERIMENTAL DETAILS

Ni(CN)$_2$·3/2H$_2$O was obtained from Alfa Aesar. The sample was dried in a vacuum oven at 200 °C for 12 h to generate Ni(CN)$_2$. Synthesis of CuNi(CN)$_4$: aqueous solutions (110 mL) of 5.3020 g (0.0220 mol) K$_2$Ni(CN)$_4$·nH$_2$O (Aldrich) and 5.4899 g (0.0296 mol) CuSO$_4$·5H$_2$O (Aldrich) were added at room temperature and stirred for 4 h. The green-blue gelatinous solid formed was filtered, repeatedly washed with distilled water, to remove the soluble by-products formed during the reaction, and allowed to air dry. The product, a fine grey powder, was further dehydrated in a tube furnace at 383 K under flowing N$_2$ for 2 h and sealed in a glass tube.

The Cu$_x$Ni$_{2-x}$(CN)$_4$ solid solution compounds were synthesized from their hydrates, as described previously [11]. Aqueous solutions of CuCl$_2$·2H$_2$O and NiCl$_2$·6H$_2$O in the correct molar ratios were quickly and simultaneously added to an aqueous solution of KCN. A pale-green gelatinous solid immediately formed. After stirring for 10 h, each powder was filtered, washed with deionized water and allowed to dry in air. The products were pale-green powders of composition $Cu_xNi_{2-x}(CN)_4 \cdot 6H_2O$. These compounds were dehydrated by heating under vacuum at 473 K for 6 h to form orange-brown anhydrous $Cu_xNi_{2-x}(CN)_4$. The polycrystalline samples were mounted on a thin glass fibre using cyanoacrylate adhesive in a goniometer and powder x-ray diffraction patterns collected over the temperature range 90-473 K using a Rigaku Synergy single-crystal x-ray diffractometer (CuKα radiation).

Inelastic neutron scattering (INS) measurements were performed on ~ 3-g sample of $Cu_xNi_{2-x}(CN)_4$, using the cold-neutron, time-of-flight, time-focusing, IN6 spectrometer (Institut Laue-Langevin, Grenoble), operating in the high-resolution mode, and offering a good signal-to-noise ratio. The IN6 spectrometer supplies a typical flux of $10^6$ n cm$^{-2}$ s$^{-1}$ on the sample, with a beam-size cross section of 3×5 cm$^2$ at the sample position. The sample was placed inside a thin-walled aluminium container and fixed to the tip of the sample stick of an orange cryofurnace. An optimized small sample thickness of 2 mm was used, to minimize effects such as multiple scattering and absorption. An incident wavelength of 4.14 Å was used, offering an elastic energy resolution of 0.17 meV, as determined from a standard vanadium sample. The vanadium sample was also used to calibrate the detectors and to normalise the spectra. Data were collected up to 100 meV in the up-scattering, neutron energy-gain mode, at 150, 250, 350 and 450 K. On IN6, under these conditions, the resolution function broadens with increasing neutron energy, and it can therefore be expressed as a percentage of the energy transfer. The ILL program LAMP [15] was used to carry out data reduction and treatment, including detector efficiency calibration and background subtraction. Background reduction included measuring an identical empty container in the same conditions as sample measurements. At the used shortest-available neutron wavelength on IN6, λ = 4.14 Å, the IN6 angular coverage (~ 10 – 114°) corresponds to a maximum momentum transfer of $Q$ ~ 2.6 Å$^{-1}$.

In the incoherent approximation [16], the $Q$-averaged, one-phonon [17] generalized density-of-states (GDOS), $g^{(n)}(E)$, is related to the measured dynamical structure factor, $S(Q, E)$ from INS by

$$g^{(n)}(E) = A \left\langle \frac{e^{2W(Q)}}{Q^2} \frac{E}{n(E,T) + \frac{1}{2} \pm \frac{1}{2}} S(Q,E) \right\rangle$$

where $A$ is a normalization constant, $2W(Q)$ is the Debye-Waller factor and $n(E,T)$ is the thermal occupation factor (Bose-factor correction) equal to $[\exp(E/k_BT)-1]^{-1}$. The + or - signs correspond to neutron energy loss or gain respectively and the bra-kets indicate an average over all $Q$. It is worth noting that the Bose factor corrected dynamical structure factor $S(Q, E)$ is generally termed $\chi''(Q, E)$, and referred to as dynamical susceptibility.

COMPUTATIONAL DETAILS

*Ab initio* density functional theory calculations were performed using the projector-augmented wave potentials [18,19] with the Vienna *ab initio* simulation package (VASP) [20–23]. The generalized gradient approximation was adopted using the Perdew-Burke-Ernzerhof (PBE) density functional scheme [24,25]. In order to account for possible weak interactions between the layers, different methods for including van der Waals' corrections were considered. These included the DFT-D2, DFT-TS, DFT-D3 and DFT-D3 (BJ) Grimme-type corrections [26–29] and the DFT-based van der Waals' scheme, vdW-DF2 [30–33]. The valence electronic configurations of Cu, Ni, N and C as used for pseudopotential generation were $3d^{10}4s^1$, $3d^94s^1$, $2s^22p^3$ and $2s^22p^2$, respectively. Ionic relaxations of $CuNi(CN)_4$ were carried out using a *k*-point mesh generated using the Monkhorst-Pack method [34] appropriate for each structural model, until the free-energy change between two ionic steps was less than $10^{-6}$ eV. Residual forces on the metals and non-metals remained below 0.03 and 0.002 eV/atom, respectively. Gaussian broadening was implemented with a smearing width of 0.01 eV.

Phonon calculations using the direct method [35] were performed on a $2 \times 2 \times 2$ supercell. The calculated partial vibrational density-of-states (PDOS), $g_k(E)$, of the $k^{th}$ atom is related to the GDOS via

$$g^{(n)}(E) = B \sum_k \left(\frac{4\pi b_k^2}{m_k}\right) g_k(E)$$

where $B$ is a normalization constant, $b_k$ is the neutron scattering length and $m_k$ the mass of the $k^{th}$ atom [36]. The constant $(4\pi b_k^2/m_k)$ represents the atom's neutron weighting factor.

The NTE coefficient was calculated using the quasi-harmonic approximation (QHA). This method uses phonon calculations for different values of the *a* lattice parameter to extract the isothermal mode Grüneisen parameters, $\gamma_i^{a,T}$, defined as,

$$\gamma_i^{a,T} = \left(\frac{d \ln \omega_i}{d \ln a}\right)_T$$

where $\omega_i$ is the mode frequency [37]. All subsequent mentions of the calculated Grüneisen parameters in this paper are referring to this definition. The QHA neglects explicit anharmonic effects, which is valid so long as these do not dominate the total anharmonicity [38]. The phonon density-of-states, mode frequencies and Grüneisen parameters were determined using the Phonopy software [39].

RESULTS AND DISCUSSION

Figure 2 shows the temperature evolution of the measured Bose-factor-corrected dynamical structure function, $S(Q,E)$, of $CuNi(CN)_4$ compared to that of $Ni(CN)_2$. Phonon dispersions emanating from Bragg spots at 2.0 and 2.6 Å$^{-1}$ along the elastic line can be seen in both, albeit much clearer in $Ni(CN)_2$. This difference is partly affected by the smaller scattering length of Cu compared to Ni, but also mirrors the sharpness of the corresponding Bragg peaks seen in the x-ray diffractograms of both compounds [11]. The Bragg spot at 2.6 Å$^{-1}$ is the (200) reflection in $Ni(CN)_2$ and (220) reflection in $CuNi(CN)_4$. The diffraction at 2.0 Å$^{-1}$ represents an (00*l*) reflection in both compounds and hence its broader nature in $CuNi(CN)_4$ reflects the greater disorder and smaller crystallite size in the *c* direction than for $Ni(CN)_2$ [11]. A secondary phonon band at ~16 meV in Ni(CN)$_2$ is shifted to lower energy in CuNi(CN)$_4$, indicating certain modes have lower frequency in CuNi(CN)$_4$. Previous magnetic measurements were carried out for CuNi(CN)$_4$, which revealed no magnetic transition down to 1.8 K [11]. This also is confirmed by our present INS measurements, which do not show any detectable magnetic feature within the above described experimental conditions.

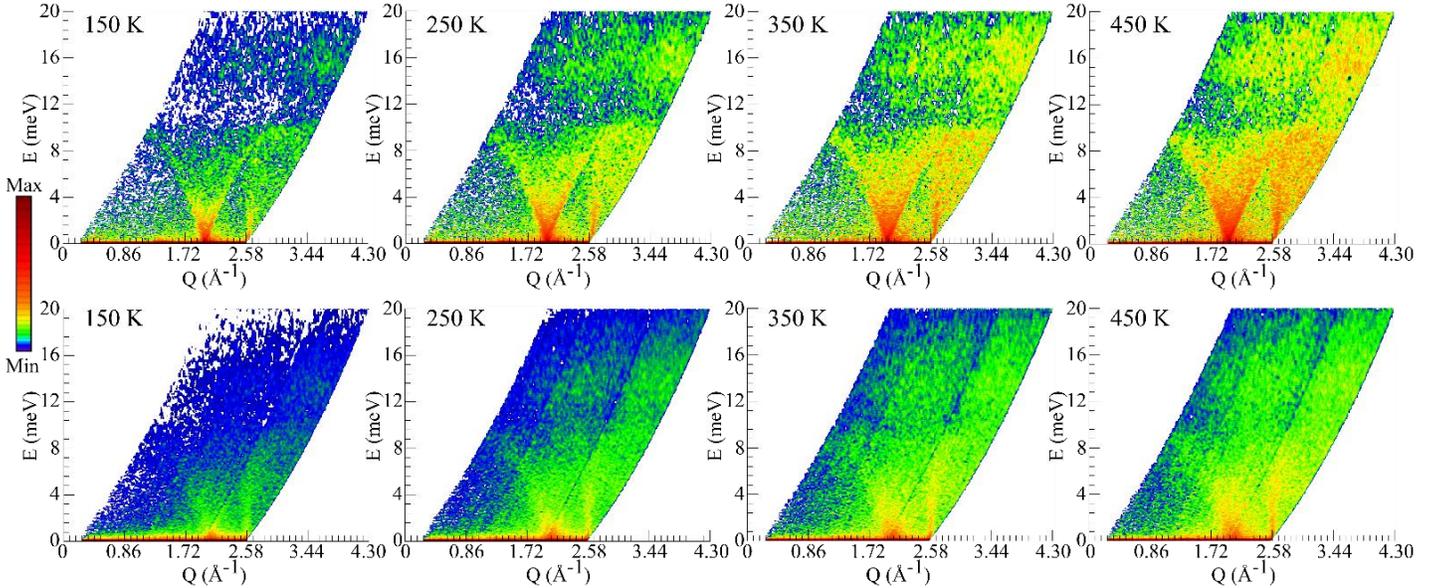

Figure 2: The temperature evolution of the Bose-factor-corrected dynamical structure function $S(Q,E)$ of Ni(CN)$_2$ (top) and CuNi(CN)$_4$ (bottom), determined from our INS measurements.

The temperature-dependent phonon spectra of CuNi(CN)$_4$ and Ni(CN)$_2$ are shown in Figure 3. The first band in both the CuNi(CN)$_4$ and Ni(CN)$_2$ spectra is observed at ~ 9.5 meV. In the Ni(CN)$_2$ spectra, the band is sharp and well defined at all temperatures, however in the CuNi(CN)$_4$ spectra, it is a rounded shoulder to the second band. For both compounds, the intensity of a sharp feature at ~ 22 meV diminishes with temperature, disappearing completely in the case of CuNi(CN)$_4$. Around 36 meV a new peak appears on increasing the temperature in the CuNi(CN)$_4$ spectra, which is absent in Ni(CN)$_2$. A phonon growth and collapse such as this is presumably the result of changes in symmetry. In CuNi(CN)$_4$, we speculate that the change is probably occurring in the stacking order, due to the layers sliding over one another more easily as temperature and interlayer separation increases. These changes were not detected in diffraction experiments [11], and were likely masked by the high level of stacking disorder inherent in this material.

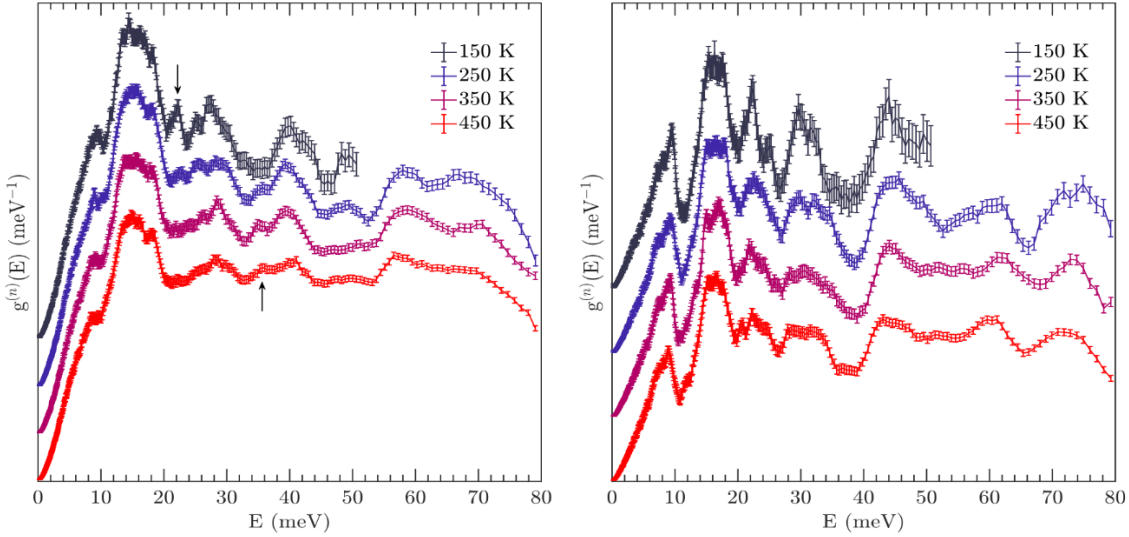

Figure 3: The temperature evolution of the phonon spectra of CuNi(CN)$_4$ (left) and Ni(CN)$_2$ (right), determined from our INS measurements. The phonon collapse and growth in CuNi(CN)$_4$ are highlighted with arrows.

The linear thermal expansion coefficients of the solid-solution compounds, Cu$_x$Ni$_{2-x}$(CN)$_4$ ($x = 0.5$, 0.33, 0.25. 0.1), extracted from our variable temperature x-ray diffraction (XRD) measurements, are shown in Table 1, together with the corresponding values for Ni(CN)$_2$ ($x=0$) and CuNi(CN)$_4$ ($x=1$). The results show that all these compounds exhibit 2D NTE behavior, with, interestingly, CuNi(CN)$_4$ showing the strongest NTE behavior.

Table 1: Measured linear thermal expansion coefficients of Cu$_x$Ni$_{2-x}$(CN)$_4$ over the temperature range 90–473 K.

| $x$ in Cu$_x$Ni$_{2-x}$(CN)$_4$ | $\alpha_a \times 10^{-6}$ K$^{-1}$ | $\alpha_c \times 10^{-6}$ K$^{-1}$ |
|---|---|---|
| 0[a,c] | -6.5 (1) | +69 (1) |
| 0.1[b] | -7.5 (4) | +81 (1) |
| 0.25[b] | -7.8 (5) | +79 (1) |
| 0.33[b] | -7.5 (5) | +78 (1) |
| 0.5[b] | -3.3 (1) | +70 (1) |
| 1[a] | -9.7 (8) | +89 (1) |

a) From reference [11]. b) This work. c) Reported values for Ni(CN)$_2$ ($x = 0$) are for a 12–295 K range.

The measured scattering functions $S(Q,E)$ of Cu$_x$Ni$_{2-x}$(CN)$_4$ ($x = 0.5, 0.33, 0.25. 0.1$) at 350 K are shown in Figure 4. They all appear to resemble that of CuNi(CN)$_4$ more closely than that of Ni(CN)$_2$, indicating that Cu$^{2+}$ has a noticeable effect on the dynamics of the compound, even at low concentrations.

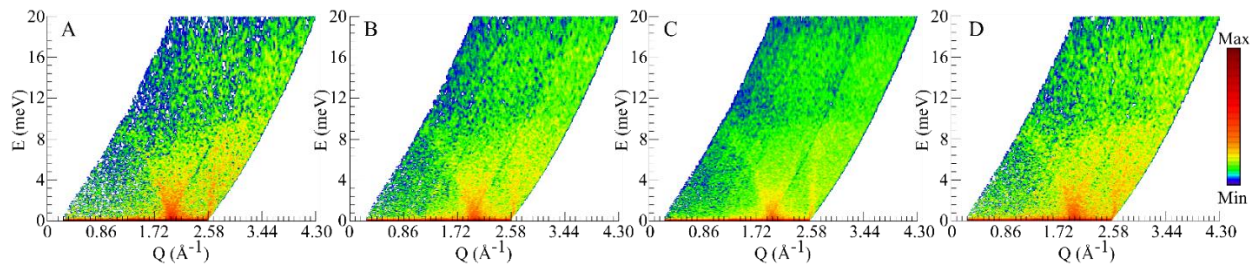

Figure 4: The Bose-factor-corrected dynamical structure function $S(Q,E)$ at 350 K of $Cu_xNi_{2-x}(CN)_4$ with $x = 0.5$ (A), 0.33 (B), 0.25 (C), 0.1 (D), determined from our INS measurements.

The phonon spectra at 350 K of $Cu_xNi_{2-x}(CN)_4$ ($x = 1, 0.5, 0.33, 0.25, 0.1, 0$) are shown in Figure 5. A clear trend can be seen as most phonon bands shift to lower energy with increasing Cu content, which is illustrated on the right-hand side of the figure. The greatest shift is observed in band 3, at ~ 44 meV. Phonon calculations discussed later showed that this band in $CuNi(CN)_4$ corresponds to localized librational motions of the CN ligand, both in and out-of-plane, which naturally leads to in-plane NTE. Hence as the $Cu^{2+}$ fraction increases and these motions become easier, a greater thermal response is induced in $CuNi(CN)_4$ compared to $Ni(CN)_2$.

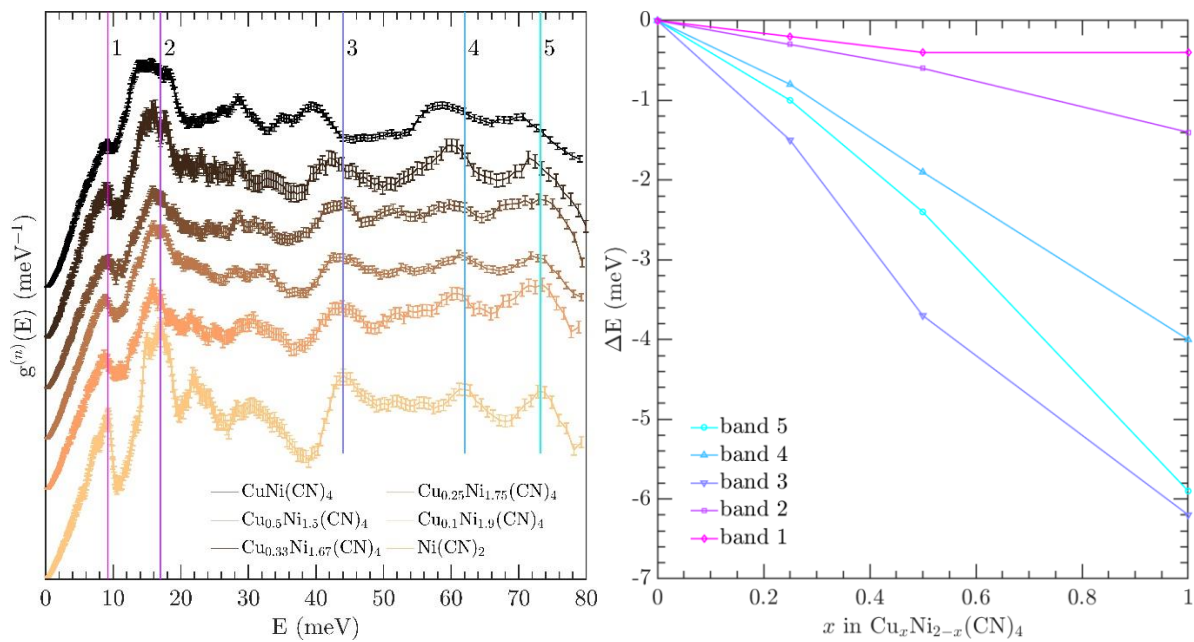

Figure 5: Left: The evolution of the phonon spectra, at 350 K, of the $Cu_xNi_{2-x}(CN)_4$ compounds as a function of the Cu content, $x$, determined from our INS measurements. As the Cu content increases, the bands marked by solid vertical lines for clarity at ~ 10, 17, 44, 62 and 73 meV all clearly shift to lower energy. The first band at 10 meV also broadens significantly. The band at 22 meV is distinct in $Ni(CN)_2$ ($x = 0$), but not present in the Cu-Ni compounds ($x \neq 0$). Right: Frequency shift of the various phonon bands, identified on the left panel, with increasing Cu content. All phonon bands shift to lower energy, with band 3, at ~ 44 meV, shifting the most. This band corresponds to localized librational motions of the CN ligand, both in and out-of-plane, and contributes to the NTE (see Figure 12). The bands 4 and 5 are from single-bond stretches.

Both Ni(CN)$_2$ and CuNi(CN)$_4$ are layered compounds with disordered stacking, and their structures were subject to previous studies [10] and [11]. CuNi(CN)$_4$ is an unusual example of an extended solid containing Cu (II) d$^9$ with a square-planar configuration. Total neutron diffraction studies of Ni(CN)$_2$ and CuNi(CN)$_4$ confirm that the layers are planar in each case [10,11], and room-temperature powder x-ray diffraction measurements show an average interlayer separation of 3.20 and 3.09 Å, respectively. Although there is no long-range order perpendicular to the layers, there is a preferred local stacking pattern. Chippindale *et al.* used neutron total scattering data to extract the pair correlation function of the compound and compare it to that of different structural models [11]. Of the potential models considered, the one that best fits the total neutron scattering data is labeled as Model 3 in Figure 6. Bond lengths and crystallographic information can be found in the supporting information of reference [11].

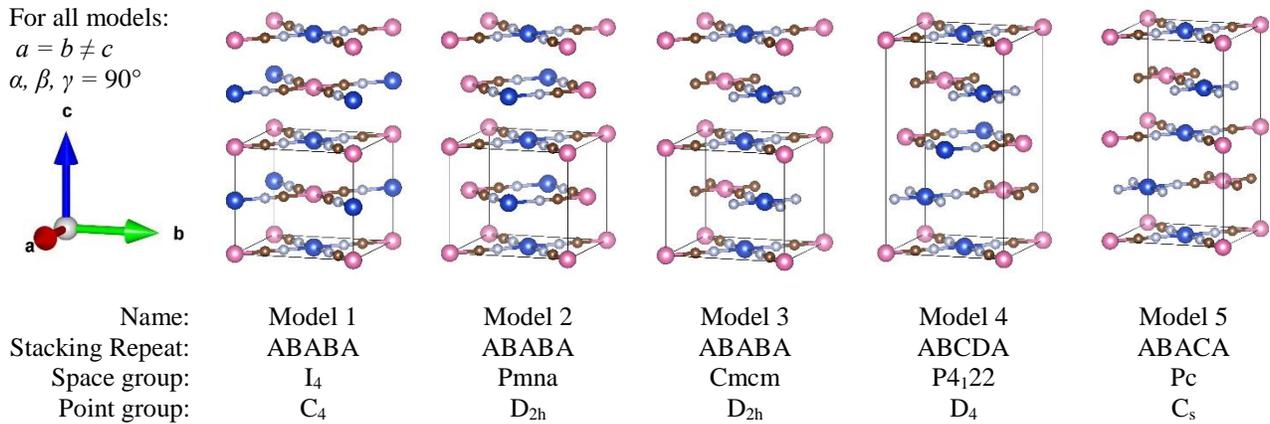

| | Model 1 | Model 2 | Model 3 | Model 4 | Model 5 |
|---|---|---|---|---|---|
| Name: | | | | | |
| Stacking Repeat: | ABABA | ABABA | ABABA | ABCDA | ABACA |
| Space group: | I4 | Pmna | Cmcm | P4$_1$22 | Pc |
| Point group: | C$_4$ | D$_{2h}$ | D$_{2h}$ | D$_4$ | C$_s$ |

For all models:
$a = b \neq c$
$\alpha, \beta, \gamma = 90°$

Figure 6: Schematic illustrations of the different structural models for the local stacking patterns of CuNi(CN)$_4$. Key: Cu, blue, Ni, pink, C, brown, N, grey. The letters under each model represent the stacking sequence. Model 3 best reproduced the measured pair correlation function from total neutron scattering measurements [11]. All the models are based on the bond lengths of Model 3, whose structural information can be found in the supporting information of reference [11].

Initial calculations were carried out on Model 3, by neglecting weak interactions. Attempts at a full ionic relaxation for this system resulted in an overexpansion of the *c* lattice parameter and an unphysical interlayer separation greater than the experimental value [11]. All further calculations using this approach were therefore carried out on partially relaxed structures with a fixed *c* lattice parameter of 6.16 Å. This result highlighted the repulsive electrostatic behavior between the layers and the importance of the weak dispersive interactions stabilizing the structure. The application of a van der Waals' (vdW) correction, in terms of different schemes incorporated into DFT [26–29], to account for weak interactions between the layers, allowed for total ionic relaxation with physically meaningful results.

In the DFT-D2 method a single $f(r_{ij})c_{ij}/r_{ij}^6$ term is used, where $c_{ij}$ is the dispersion coefficient, $r_{ij}$ is the distance between atom *i* and *j* and $f(r_{ij})$ is a Fermi-type damping function [26]. The dispersion coefficients and damping function between different atoms pairs are calculated using predefined values of atomic vdW radii and dispersion coefficients. This method resulted in a *c* lattice parameter just below the experimental value at 15 K. In the Tkatchenko-Scheffler variation of this method (DFT-TS) [27], the dispersion coefficients and vdW radii are given a charge density dependence, which led to an under correction of the attractive forces and hence a value of $c$ larger than experiment. When instead, a second $c_{ij}/r_{ij}^8$ term is added (DFT-D3) [28], the outcome was even less accurate. However, in DFT-D3, an alternative Becke-Jonson damping function can be used, labeled as DFT-D3 (BJ) [29], which resulted in lattice parameters that were the closest to those experimentally measured [11]. Although using these corrections allowed for total ionic relaxations, their use resulted in some imaginary phonon modes pointing towards possible structural instabilities, which are discussed later.

In a further refinement of the process, the vdW-DF2 function, which includes vdW corrections within the exchange-correlation functional, was used, providing a more accurate approach of including weak interactions [30–33]. This method predicts a reasonable experimental $c$ parameter, however it overestimates the $a$ lattice parameter. Figure 7 shows the outcome of the structural relaxation using different DFT schemes, along with experimentally determined values.

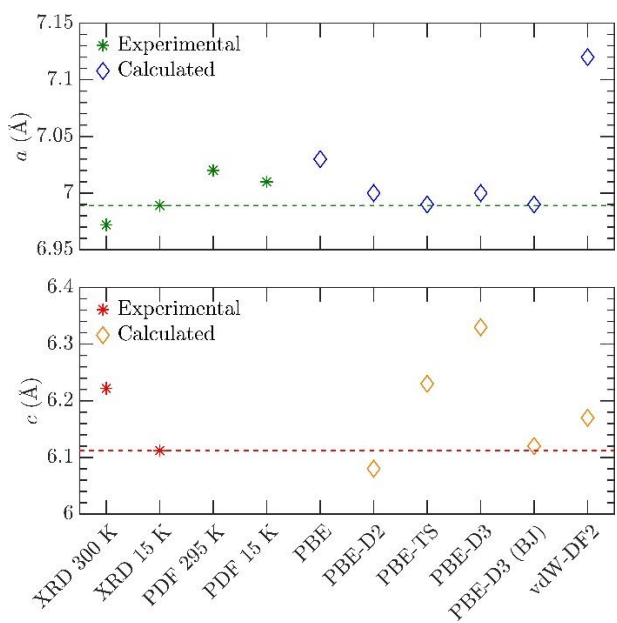

Figure 7: Upper panel and lower panel show the $a$ and $c$ lattice parameters, respectively, from the structural relaxation of Model 3 of CuNi(CN)$_4$, using different DFT schemes. The experimental values are from x-ray diffraction (XRD at 15 K and 300 K) and neutron pair distribution function analysis (PDF 15 K and 295 K) [11]. Note that there is no $c$ lattice parameter derived from the experimental neutron total scattering PDF measurements of the layered CuNi(CN)$_4$. The $a$ parameter derived from the same PDF data is only approximate, and represents the sum of the average interatomic distances and hence will always be larger than the actual lattice parameter, and will increase with temperature regardless of NTE. There is no $c$ parameter for the structural relaxation using the PBE DFT functional, without a vdW correction, as its value was ~7.1 Å, and hence off the scale visible here. The dashed lines mark the experimental parameters from XRD measurements at 15 K, and are guide for the eyes to help compare how the DFT relaxed parameters (at 0 K) agree with or deviate from the measurements, following the adopted computational scheme.

Having established the effect of different DFT computational schemes on Model 3, similar calculations were carried out on the other model structures (Figure 6) for the sake of comparison. Model 1 and Model 5 were difficult to converge to a reasonable ionic ground state. This indicates instabilities in their structure, and hence does not represent any local order in CuNi(CN)$_4$, which is in agreement with the observations [11]. Regardless of which DFT-based approximations were used, the structure of Model 4 converged to the lowest ground-state energy. The difference in energy between Model 4 and Model 3 was in all cases very small, approximately 0.01%. The lack of a clear energetically favorable stacking pattern provides a further explanation for the disorder. Ultimately, Model 3, rather than Model 4 was used for the present lattice dynamical study as Model 3 is experimentally refined and has a smaller unit cell, and hence has the advantage of being less computationally demanding. Furthermore, trial phonon calculations carried out using Model 4 gave very similar results to those obtained from the calculations using Model 3.

To further validate the structural model for the phonon study, the electronic band gap was calculated for Model 3, as well as for Ni(CN)$_2$. For the latter, the structural model published in [10] was used. The experimental band gaps extracted from diffuse reflectance spectra [11] are 2.0 and 2.7 eV for Ni(CN)$_2$ and CuNi(CN)$_4$, respectively, revealing that adding Cu$^{2+}$ to Ni(CN)$_2$ leads to an increase in the electronic band gap. The calculated band gaps reproduce well the observed values, with estimated values of 1.95 and 2.59 eV for Ni(CN)$_2$ and CuNi(CN)$_4$, respectively.

The generalized density of states (GDOS) of Model 3, calculated using different DFT schemes, with and without a vdW correction, are shown in Figure 8. The importance of including a vdW correction is clear when comparing results from the PBE and PBE-D2 calculations. Results from PBE-D2 calculations match the measured data significantly better than the former, especially at low energy. The third band ~ 20 meV, in the PBE-D2 GDOS is shifted to higher-energy compared to the other calculated spectra, reproducing a feature that can be clearly seen in the measurement at 450 K. The atom-resolved neutron-weighted partial densities of states (PDOS) are also shown in Figure 8. Here the effect of the different Cu and Ni scattering lengths can clearly be seen. The cyanide stretching region is also shown. The calculated stretching energies agree well with IR (270.4 meV) and Raman (270.8, 273.9) data [11].

Though the vdW-DF2 functional does take into account weak interactions and it correctly predicts the experimental *c* lattice parameter, it does not give a good overall description of the dynamics of the system, as its corresponding GDOS does not compare well with the measured spectra. Furthermore, above 50 meV, vibrational bands from the vdW-DF2 calculation are shifted to lower energy. As these modes involve single-bond stretches, this shift is due to the increased *a* lattice parameter resulting from this functional. Although the PBE-D2 method best reproduced the measured GDOS, it resulted in a tiny imaginary phonon feature, also present with all other Grimme-type corrections trialed. This suggests that including vdW interactions leads to some instabilities within the ordered structure, which must then be mitigated by the stacking disorder in the real structure. Interestingly, the calculations reproduce well the high-temperature spectra, but do not predict the low-temperature phonon band observed at ~22 meV, which experimentally collapses with increasing temperature. The calculations also reproduce the feature around 35 meV, which only appears experimentally above 250 K. Hence, the evolution with temperature, revealed by the INS measurements, results in a greater similarity to the ordered structural model at higher temperatures. These changes can tentatively be interpreted as likely due to the fact that when the layers in CuNi(CN)$_4$ move apart (due to positive thermal expansion along the *c* axis), they can slide over one another more easily and the interaction leading to the disorder between them decreases.

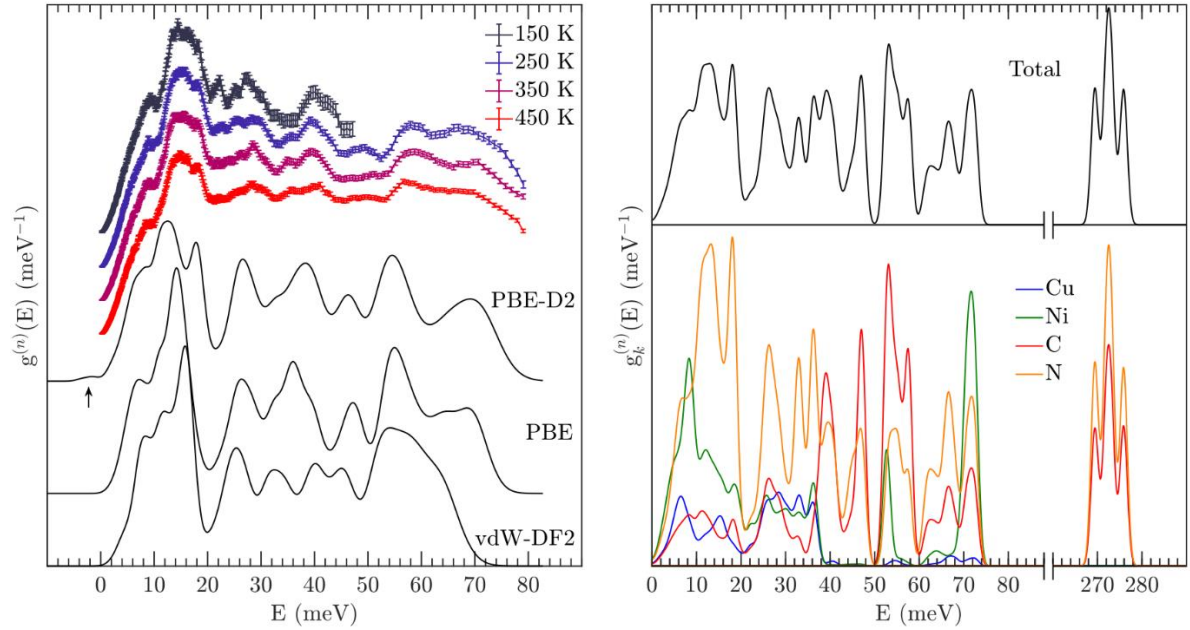

Figure 8: Left: Comparison of the measured and calculated GDOS of CuNi(CN)$_4$. Calculations using Model 3 were performed using different DFT-based approaches (0 K), where weak interactions were considered (PBE-D2 and vdW-DF2) or neglected (PBE). The PBE-D2 scheme best reproduced the measured GDOS, although it resulted in a tiny imaginary feature, indicated by the arrow, and points towards some phonon instabilities (illustrated in Figure 10). The calculated spectra have been convoluted with a Gaussian of FWHM of 10% of the energy transfer to mimic the experimental instrumental resolution. Right: The neutron-weighted PDOS for each atom type calculated using the PBE-D2 scheme. The total GDOS is obtained by summing the partial contributions, taking into account the stoichiometry.

The calculated dispersion curves of Model 3 of CuNi(CN)$_4$, with and without vdW correction, are shown in Figure 9, along with the Brillouin zone high-symmetry points. There is the clear similarity in the dispersion curves along the Γ-X-S-Y-Γ path and the Z-U-R-T-Z path. This is expected, as both paths are perpendicular to *c**. However, many degenerate modes on the Z-U-R-T-Z path are split along Γ-X-S-Y-Γ. The latter path involves simultaneous vibrations of both layers, whereas the former corresponds to motions of a single layer, whilst the other is at rest. The inclusion of a vdW correction (Figure 9) results in the acoustic modes becoming unstable at the Γ and Z high-symmetry points.

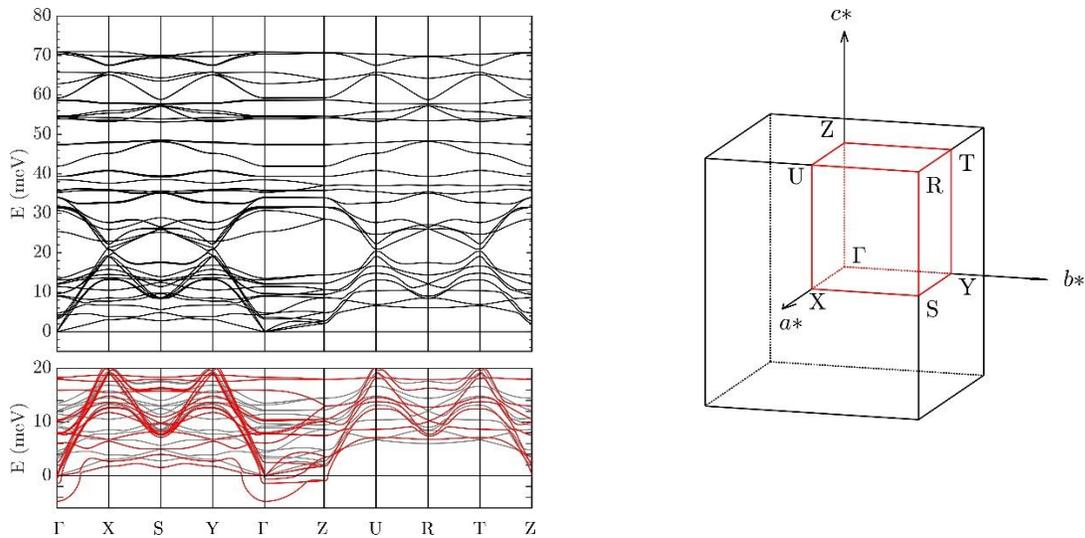

Figure 9: Left: The dispersion curves of CuNi(CN)$_4$ Model 3, calculated using the PBE functional with no vdW correction (top), and using a Grimme-type vdW correction (bottom). The unstable phonon modes can clearly be seen at the Γ and Z high-symmetry points. Right: The high-symmetry points in the first Brillouin zone of Model 3 of CuNi(CN)$_4$.

The eigenvectors of the two most unstable phonon modes at the Γ-point resulting from the vdW correction are shown in Figure 10. The first mode is an undulation of the layers, whilst the second corresponds to a sliding motion of the layers over one another, indicating the inadequacy of the ordered model in describing the disordered stacking arising from shear stress. Attempts at phonon calculations using the vdW-DF2 scheme, at non-equilibrium volumes, resulted in very similar unstable modes. These instabilities further confirm that it is indeed the weak interactions between the layers that drives the disorder in the material.

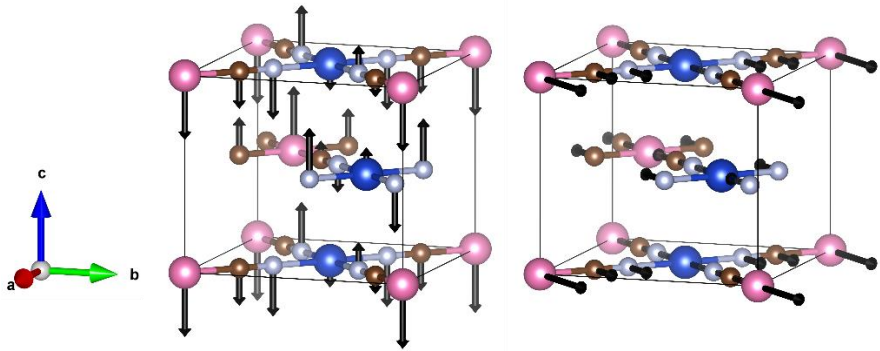

Figure 10: Schematic illustration of the unstable phonon modes in CuNi(CN)$_4$ calculated including a vdW scheme. Left: the atoms are oscillating out of plane causing an undulation of the layers. Right: the layers are sliding over one another. The arrows represent the magnified real parts of the eigenvectors. Key: Cu, blue, Ni, pink, C, brown, N, grey. The absence of an arrow indicates atoms are at rest.

Due to the phonon instabilities present in the ordered structural model when including vdW corrections, the thermal expansion behavior, $\alpha_a$, was calculated by ignoring possible weak interactions. An interlayer separation of 3.08 Å was imposed, and only the *a* lattice parameter was varied to change the volume. The results are shown in Figure 11. The negative sign was correctly reproduced, and the most negative value was found to be ~ -40 × $10^{-6}$ $K^{-1}$ at 100 K; larger than the experimental value of -9.7 × $10^{-6}$ $K^{-1}$. However, the experimental value was extracted from lattice parameters obtained from XRD patterns collected over the 100 – 550 K temperature range [11]. In this range, our calculated expansion coefficient is not constant, but its average is ~ -10 × $10^{-6}$ $K^{-1}$, which is very close to the experimental value.

The contribution of modes of energy E to the 2D NTE is depicted on Figure 11. The trend mirrors what has previously been calculated for $Ni(CN)_2$ [13]. The lowest-energy modes contribute the most, with bond-stretch modes (60 – 80 meV) acting to decrease the NTE at higher temperatures. A clear jump in the trend at 40 meV reveals modes of this energy also contribute to the NTE. Analysis of the mode eigenvectors between 40 and 50 meV, revealed that these phonons consist entirely of localized librational motions of the CN ligand, an example of which is shown in Figure 12. These motions are consistent with NTE mechanisms within the layers, as they are a classic example of the tension effect [9]. In the phonon spectra (Figure 3), the band at 40 meV in $CuNi(CN)_4$ is shifted to a lower energy with respect to $Ni(CN)_2$. Hence, although the interlayer separation is lower in $CuNi(CN)_4$ than $Ni(CN)_2$, the out-of-plane motions require less energy in $CuNi(CN)_4$.

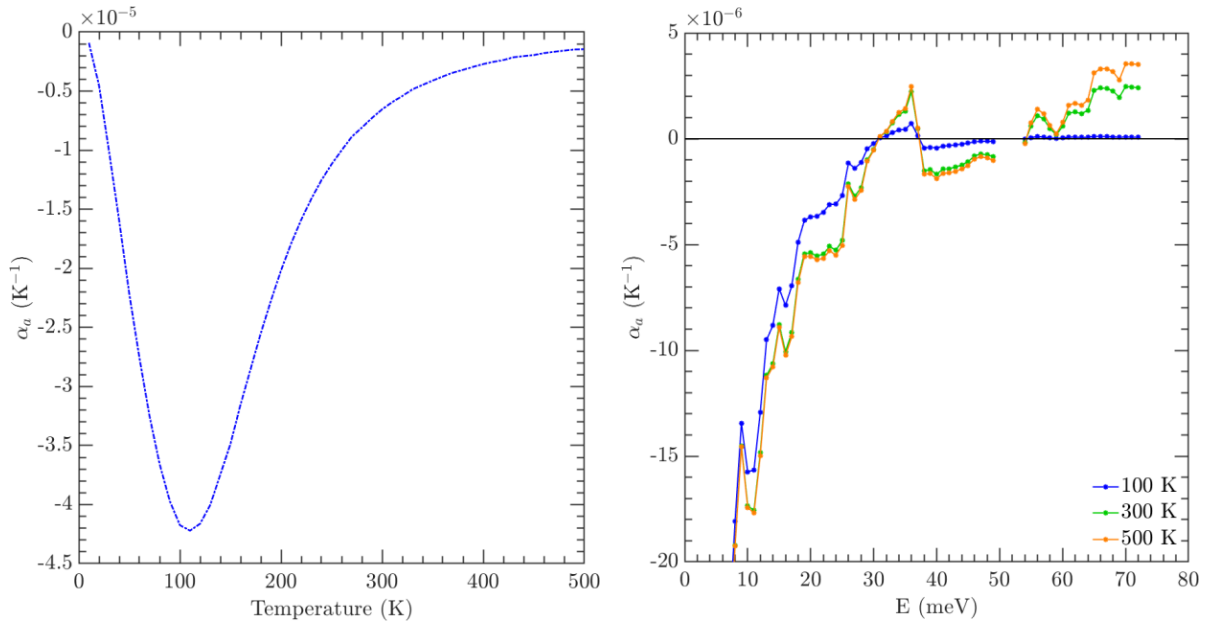

Figure 11: Left: The calculated linear thermal expansion coefficient, $α_a$, of $CuNi(CN)_4$. Right: Contribution to the negative thermal expansion of modes of energy E at three different temperatures in $CuNi(CN)_4$.

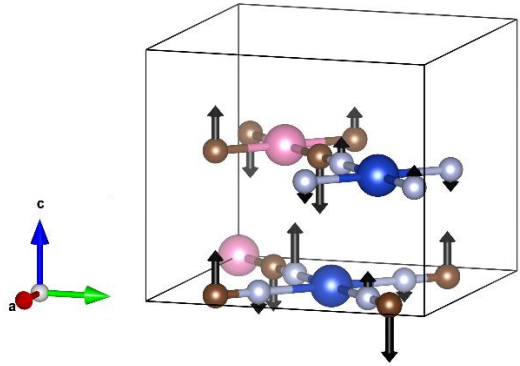

Figure 12: Schematic illustration exemplifying a Γ point mode at ~40 meV in CuNi(CN)$_4$, consisting of a localized librational motion of the CN ligand. Note the metals, without arrows, are at rest. The arrows represent the real displacement vectors of the atoms, with a magnified amplitude. Key: Cu, blue, Ni, pink, C, brown, N, grey.

The adequacy of the QHA used in this work can further be validated by calculating the implicit anharmonicity, arising from the volume effect, and comparing it to the total anharmonicity (both volume and temperature effects) derived from the measured GDOS, which includes also the explicit anharmonic component not described within the framework of the QHA (Figure 13). The total anharmonicity, which is equal to ($d\ln E_i/dT$) was extracted from the cumulative GDOS measured at 150 and 450 K. The implicit anharmonicity is equal to $-\gamma_i^T \alpha_V$, however here we can only show the contribution along $a$ ($-\gamma_i^{a,T} \alpha_a$). Figure 13 shows the anharmonicity of phonon modes of energy E, and highlights that the experimental trend is well reproduced computationally within the QHA framework. We also notice a difference in absolute value, which could likely point toward an explicit anharmonic contribution. Nonetheless, it is found that the QHA is capable of giving valuable insights into the NTE behavior.

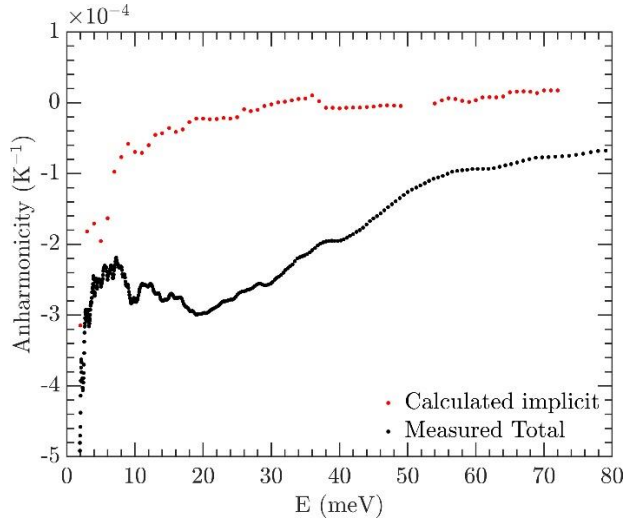

Figure 13: The experimental total anharmonicity (black) of CuNi(CN)$_4$ extracted from the cumulative GDOS at 350 and 450 K, compared with the calculated implicit anharmonicity (red) derived using the quasi-harmonic approximation. The difference between the two represents the explicit anharmonicity.

The isothermal Grüneisen parameters used to calculate the NTE are presented in Figure 14. They show how specifically modes around 5 meV have the largest NTE-inducing effect. The Grüneisen-filtered dispersion curves reveal that the NTE arises mainly from two optic modes, and one acoustic mode, along the high-symmetry points S, Γ and Z. These modes involve motions where the [NiC$_4$] units translate along $c$, but the [CuN$_4$] units deform, with N atoms along one diagonal moving in opposite directions to the N atoms along another diagonal. The extreme picture of this type of deformation would be an oscillation between a square-planar and tetrahedral shape. This type of deformation does not occur in Ni(CN)$_2$ [10,12], as a d$^8$ square-planar configuration is more energetically stable compared to the d$^9$ configuration found for Cu$^{2+}$.

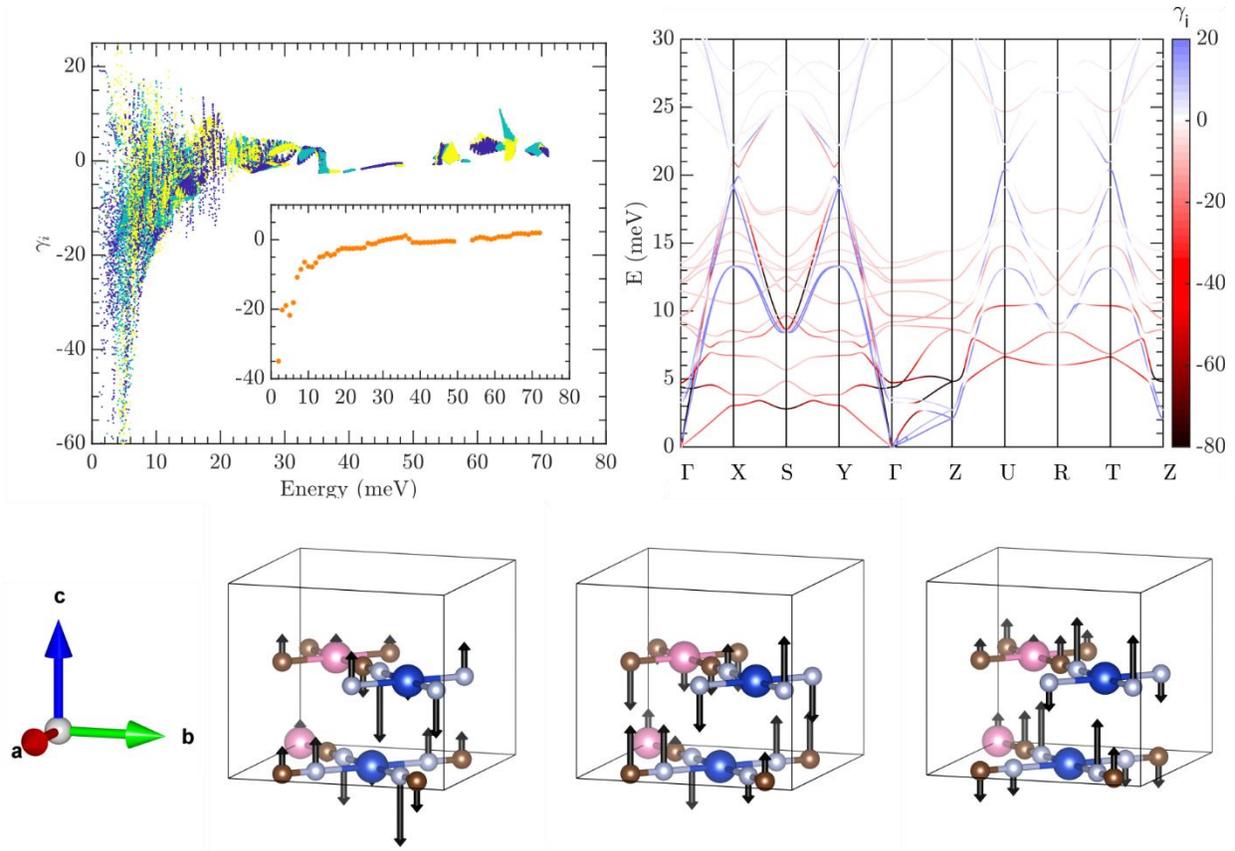

Figure 14: Top left: The calculated mode Grüneisen parameters of CuNi(CN)$_4$. Neighboring phonon modes are colored differently for contrast. Inset: The average Grüneisen parameter for modes of energy E. Top right: The dispersion curves colored according to their mode Grüneisen parameters. Bottom: Schematic illustration of the normal modes contributing the most to the NTE in CuNi(CN)$_4$. The left and central mode are the third and fourth optic modes at the $\Gamma$-point, respectively. The right-hand mode is the lowest-energy optic mode at the S-point. The lowest-energy optic mode at the Z-point, which also has a pronounced negative Grüneisen parameter, is similar to the central mode, but subject to different relative phases. The arrows represent the real displacement vectors of the atoms, with a magnified amplitude. Note the large deformation around the Cu center compared to the Ni center. Key: Cu, blue, Ni, pink, C, brown, N, grey. The absence of an arrow indicates atoms are at rest.

The deformation of the Cu center is highlighted further in the calculated thermal ellipsoids shown in Figure 15. As expected, the average displacements of all the atoms is largest along *c*. However, this displacement is greatest for N and least for Cu, consistent with the deforming motions of the [CuN$_4$] unit described in Figure 14. Another observation is the similarity of the thermal ellipsoids for Ni and C, consistent with the more rigid translations of the [NiC$_4$] units.

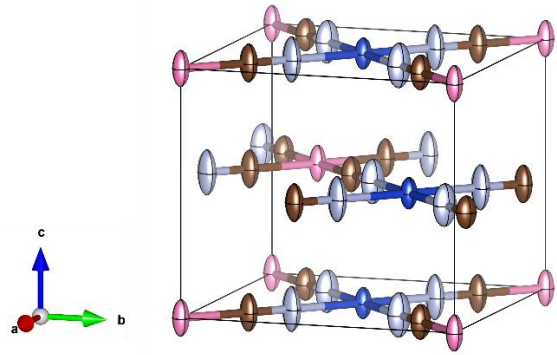

Figure 15: Schematic illustration of the calculated thermal ellipsoids of CuNi(CN)$_4$ at 300 K. The ellipsoids represent a 75% probability of containing the atom. Key: Cu, blue, Ni, pink, C, brown, N, grey.

CONCLUSIONS

The present work, combining INS measurements and *ab initio* lattice dynamical calculations, provides new insights into the anomalous thermal expansion behavior of the layered compounds, Cu$_x$Ni$_{2-x}$(CN)$_4$ ($0 \leq x \leq 1$). The emphasis is on a detailed phonon dynamics analysis of the line phase CuNi(CN)$_4$ ($x = 1$), exhibiting 2D NTE which is ~ 1.5 times larger than the value for Ni(CN)$_2$ ($x = 0$). INS measurements were analysed for the Cu$_x$Ni$_{2-x}$(CN)$_4$ solid solution for which $x \leq 0.5$. It is observed that on increasing the Cu$^{2+}$ content in the solid solution, the phonon bands shift to lower energy. The greatest shift comes from phonon modes that involve both in- and out of plane librational motions of the CN ligand. These modes induce some NTE of the *a* lattice parameter, and hence their decreased frequency likely contributes to greater NTE observed in CuNi(CN)$_4$.

Our INS measurements of CuNi(CN)$_4$ revealed a phonon collapse at 22 meV and phonon growth at 36 meV with increasing temperature, which is not observed in the spectra of Ni(CN)$_2$. This evolution in the phonon spectra points towards possible changes in stacking order of CuNi(CN)$_4$. Phonon calculations using an ordered model reproduce well the higher temperature GDOS. Hence, as temperature increases and the layers move further apart, the stacking pattern of CuNi(CN)$_4$ changes to better resemble the adopted ordered structural models.

Phonon calculations based on different DFT schemes, reveal instabilities in the ordered model stemming from undulating and slipping motions of the layers when weak interactions are included. Thus, the disordered stacking in CuNi(CN)$_4$ stems from its interlayer attractive forces, rather than a lack of them. These forces appear greater in CuNi(CN)$_4$ than in Ni(CN)$_2$ due to the smaller interlayer separation.

Analysis of mode Grüneisen parameters and eigenvectors reveal that the most NTE-inducing phonons include a large deformation of the [CuN$_4$] units, whilst the [NiC$_4$] units move as a rigid entity. This type of deformation allows for greater out-of-plane motion of the N atoms. As these motions were not found in previous studies of Ni(CN)$_2$, we believe they are the reason for the enhanced NTE in CuNi(CN)$_4$ compared to Ni(CN)$_2$. It is likely the deformation exists for the [CuN$_4$] unit, but not the [NiC$_4$] unit, due to the d$^9$ configuration of Cu$^{2+}$ compared to d$^8$ for Ni$^{2+}$. The present study opens the door for further investigations on two levels. Firstly, using electronic structure and orbital calculations could shed further light on the effect of chemical substitution (Ni$^{2+}$ by Cu$^{2+}$) on phonon dynamics and account for the thermal expansion behavior observed for CuNi(CN)$_4$. Secondly, the presently reported INS and NTE measurements of the solid solutions ($x$ = 0.1, 0.25, 0.33 and 0.5), in addition to CuNi(CN)$_4$ ($x$ = 1) and Ni(CN)$_2$ ($x$= 0), reveal interesting trends calling for a dedicated analysis taking into account the suspected non-planar layers of the solid-solution compounds.

ACKNOWLEDGEMENT

The ILL is thanked for providing beam time on the IN6 spectrometer. The use of the Chemical Analysis Facility (CAF) at Reading is acknowledged. S. d'A. thanks the ILL and the University of Reading for the PhD studentship.